\documentclass[acmsmall]{acmart}

\pdfoutput=1

\usepackage[utf8]{inputenc}
\usepackage[T1]{fontenc}

\setcopyright{rightsretained}
\copyrightyear{2020}
\acmYear{2020}
\acmConference[HATRA 2020]{HATRA 2020: Human Aspects of Types and Reasoning Assistants}{15--20 November, 2020}{Chicago, IL}
\acmBooktitle{HATRA 2020: Human Aspects of Types and Reasoning Assistants, 15--20 November, 2020, Chicago, IL}
\usepackage{listings, listings-rust}

\usepackage{geometry, marginnote}

\begin{document}

\title[Making formal normal]{Towards making formal methods normal: meeting developers where they are}

\author{Alastair Reid}
\email{adreid@google.com}
\orcid{0000-0003-4695-6668}
\affiliation{%https://www.overleaf.com/project/5f61dbf562379200014142db
    \institution{Google Research}}

\author{Luke Church}
\email{luke@church.name}
\orcid{0000-0002-9943-6597}
\affiliation{%
    \institution{Computer Laboratory, University of Cambridge}}

\author{Shaked Flur}
\email{sflur@google.com}
\orcid{0000-0002-8281-4926}
\affiliation{%
    \institution{Google Research}}

\author{Sarah de Haas}
\email{dehaass@google.com}
\orcid{0000-0001-7487-0269}
\affiliation{%
    \institution{Google Research}}

\author{Maritza Johnson}
\email{maritzaj@google.com}
\orcid{0000-0002-6305-0131}
\affiliation{%
    \institution{Google Research}}

\author{Ben Laurie}
\email{benl@google.com}
\affiliation{%
    \institution{Google Research}}

\renewcommand{\shortauthors}{Reid, et al.}

\begin{abstract}

    Formal verification of software is a bit of a niche activity: it is only
    applied to the most safety-critical or security-critical software and it is
    typically only performed by specialized verification engineers.
    This paper considers whether it would be possible to increase adoption of formal
    methods by integrating formal methods with developers' existing practices and workflows.
    We do not believe that widespread adoption will follow from making the prevailing formal
    methods argument that correctness is more important than engineering teams realize.
    Instead, our focus is on what we would need to do to enable programmers
    to make effective use of formal verification tools and techniques.
    We do this by considering how we might make verification tooling that
    both serves developers' needs and fits into their existing development lifecycle.
    We propose a target of two orders of magnitude increase in adoption within a decade
    driven by ensuring a positive `weekly cost-benefit' ratio for developer time invested.
    
\end{abstract}

\iffalse
\begin{CCSXML}
<ccs2012>
 <concept>
  <concept_id>10010520.10010553.10010562</concept_id>
  <concept_desc>Computer systems organization~Embedded systems</concept_desc>
  <concept_significance>500</concept_significance>
 </concept>
 <concept>
  <concept_id>10010520.10010575.10010755</concept_id>
  <concept_desc>Computer systems organization~Redundancy</concept_desc>
  <concept_significance>300</concept_significance>
 </concept>
 <concept>
  <concept_id>10010520.10010553.10010554</concept_id>
  <concept_desc>Computer systems organization~Robotics</concept_desc>
  <concept_significance>100</concept_significance>
 </concept>
 <concept>
  <concept_id>10003033.10003083.10003095</concept_id>
  <concept_desc>Networks~Network reliability</concept_desc>
  <concept_significance>100</concept_significance>
 </concept>
</ccs2012>
\end{CCSXML}

\ccsdesc[500]{Computer systems organization~Embedded systems}
\ccsdesc[300]{Computer systems organization~Redundancy}
\ccsdesc{Computer systems organization~Robotics}
\ccsdesc[100]{Networks~Network reliability}

\keywords{datasets, neural networks, gaze detection, text tagging}
\fi

\maketitle

\section{Introduction}

Mainstream commercial software development processes are currently beyond the
scope of many kinds of formal methods techniques that aim to increase our
assurance in the correct operation of software.
Increasing the use of formal methods is hard and expensive because of the
significant impact of applying formal methods on the software development
process.
This limits the application of formal methods to the organizations that have
a reason to value formal methods: essentially a small number of applications
that require the additional assurance of a formal correctness proof (e.g., the
usual safety-critical systems).
The question we tackle is: how can we increase the access to and adoption of
formal methods tools and techniques?

Currently researchers are making these high assurance
methods easier to adopt through
education~\cite{krishnamurthi:fm:2019},
case studies
and improvements to the tools and methodologies that make the high powered
technologies easier to approach.
In this paper, we consider the problem from the other end: how to bring
a little more assurance into routine developer practice --- making the everyday
technologies a little more powerful. The two approaches are not in opposition: both 
are needed and valuable for broadening the adoption of formal methods.
We informally estimate that significantly less than 1\% of developers use formal verification so
we suggest a useful goal would be to increase the number of developers 
using some formal methods by two orders of magnitude over the next decade.

We believe that critical to this is to ensure that the additional effort of
using formal methods tools has an obvious, short-term benefit that justifies it. 
For this approach to work, we have to keep an eye on both sides of the 
`weekly cost-benefit' ratio, making sure over each week of using the tool 
that the additional cost always pays for itself in benefit it returns. 

Our core idea of how to achieve this is to meet developers where they are: building 
on their existing skills and workflows, using the techniques they are familiar with, and
taking advantage of the artifacts that they currently create --- but applying them 
for new purposes. 

\section{Meeting the developer where they are}

One thing developers do a lot of is testing.
Formal methods practitioners sometimes criticize testing because it does not
guarantee the absence of bugs, but we think it is more useful to focus on what
testing and verification have in common and how they can strengthen/benefit
each other than to focus on how or when verification is superior to testing.

Testing is an important complement to verification: in verification, it is easy
to start with incorrect assumptions about the environment (e.g., network
protocols, hardware, OSes, programming language)~\cite{fonseca:ecs:2017}; while,
in testing, incorrect assumptions will usually make themselves known.

By drawing parallels between testing activities and formal verification, we can
reuse tests with formal verification tools.
This approach is rooted in Property-based testing~\cite{claessen:icfp:2000} and
Parameterized unit tests~\cite{tillmann:fse:2005}.
We are currently looking at whether a similar parallel can be made between
contract-based verification and techniques used in unit testing.

% We think that we have more to gain by looking at what testing and verification
% have in common than by looking at how they differ.

\subsection{Non-goals}

If we are to succeed in dramatically increasing the adoption of formal methods
by software engineers, we think that we will have to sacrifice a few things.

First, we will have to let go of a big goal: it is unrealistic to think that 
we will verify that the software is 100\% correct.
However, we still aim make it more 
trustworthy, more robust, etc. (Bearing in mind that 
even software with mechanized proofs of full functional correctness 
can contain bugs~\cite{fonseca:ecs:2017}.)

Second, we will have to let go of some ancillary benefits. One of the major costs 
in formal verification of realistic, real-world systems is the size and complexity 
of their specifications and the work of updating the specification as the software develops.
Writing and maintaining an accurate, thorough specification of a software
system is extremely rare.
We believe that this is because software engineering teams often do not get much
value in return for the effort of creating and maintaining a
detailed specifications.\footnote{Arguably, a thorough testing regime results
in a specification of sorts --- if only we can adapt the tests for use in
verification.}

If we start by building on parallels with testing and other standard
engineering practices, this may limit what we do with formal methods, we may lose some of
the benefits of a more narrow adoption of standard formal methods
orthodoxy. One of these benefits that formal methods practitioners often 
claim is that the act of creating a formal specification and
of focusing early on a design that makes it easy to prove the correctness of the 
system increases the quality of the software that they build. 
In the Cognitive Dimensions of Notations Framework, often used for describing trade-offs
in the design of programming languages~\cite{green:pandc:1990} this would be described as a Useful Awkwardness. 
There might be extra hoops to jump through, but the act of doing it is useful. If we avoid that 
Awkwardness, we may lose that Usefulness.
That said, practitioners of Test Driven Design would
claim that their methodology clarifies the design and testability of
their software --- so it is possible that the loss will not be too great?

We want to emphasize that our focus is on using testing as a bridge to
adoption of formal methods.
There has been a lot of work on using formal tools to assist testing such as
using symbolic/concolic execution to generate high-coverage
testcases~\cite{godefroid:cacm:2020, cadar:cacm:2013, garg:icse:2013, dimjasevic:ifm:2018}
and to improve fuzzing~\cite{beyer:hvc:2017,
visser:tacas:2020}.
While these are similar at a technical level, the different focus raises different
usability issues.

With our focus on the weekly cost-benefit ratio, we want to be very careful
about what we give up on the developer side but we will ask them to do some
additional work.
The strength of formal methods over testing is their generality: a passing test
tells you that your system works in one specific case while formal verification
tells you that your system works for a whole class of cases.
To get this benefit, developers will have to write parameterized tests that can
be applied with a range of different inputs.
This will take additional work and training/practice although we believe
that the increasing use of structure-aware fuzz-testing and the increasing
sophistication of fuzz-driver generation~\cite{babic:fse:2019} will also drive
a push towards writing more general tests.

%
% Our focus though is on how to use the existing comfort, investment, etc.  in testing to enable formal verification.

\section{Our Rust prototype}

At present, we are choosing to focus our attention on Rust.
We made this choice for a combination of technical and cultural and community reasons.
Technically, Rust's ownership type system provides
useful guarantees about Rust programs that assist formal reasoning about those
programs~\cite{astrauskas:oopsla:2019} and is designed to rule out certain
classes of bugs.
Culturally,
these technical promises of safer code
are a major motivation for teams adopting Rust
so Rust programmers are already
self-selected as people who care more about safety.
The Rust developers have put a lot of work into instilling these values into
the Rust community:
the standard Rust beginner's
book~\cite{klabnik:book:2018} focuses on starting new Rust programmers off with
a number of good habits such as creating small, testable units of code,
building testsuites, etc.

Our focus on Rust is not without its costs though.
Rust verification tools are far behind the state of C verification tools and
there is much to do in creating tools that can verify anything about Rust programs.
On the positive side,
the immaturity of Rust verification means that it is potentially possible to influence the direction of the
language's future development
and it is generally easier to adopt new practices in new code than to attempt to retrofit
them in established codebases.

Concretely, we are currently adapting existing verification tools used with
C for use with Rust and creating Rust verification
libraries~\cite{rust-verification-tools}.
We are also creating wrapper scripts around the tools to prototype our ideas
for a suitable user interface.

As a simple example of how our library furthers our goal,
consider the following simple
test expressed using the proptest fuzzing
library~\cite{proptest}.

\begin{lstlisting}[language=Rust, numbers=left, basicstyle=\ttfamily]
proptest! {
    #[test]
    fn multiply(a in 1..=1000u32, b in 1..=1000u32) {
        let r = a*b;
        assert!(1 <= r && r <= 1000000);
    }
}
\end{lstlisting}

Those familiar with Rust will recognize most of the syntax except for the
syntax ‘a in ...’ on line 3 that is used to constrain the values that ‘a’ can
take in the test and the proptest! macro that implements this syntax by
generating random numbers satisfying the constraint, invokes the test some
number of times, catches errors and reports results.
Constraints can be ranges of numbers, regular expressions, data types like
Option and BTreeMap and can be defined by users, combined and transformed to
provide an expressive, extensible domain-specific language of constraints.

With some largely superficial differences, this basic shape of interface is
supported by a variety of fuzzers (with interfaces of varying expressive power
in the constraints) and is familiar to Rust programmers.
Our verification-oriented implementation of the proptest API capitalizes on
this familiarity.
Instead of interpreting `a in ...' as random number generation, our
implementation interprets it as universal quantification.
Using our tools to verify this test checks that the program will not fail the
assertion (or raise any other assertion) for any values of `a' and `b'
satisfying the constraints.

At present, we support three ways of checking these properties.
The first is to use the fuzz-tester built into the original proptest library:
randomly sampling a subset of the input space to create a probabilistic
argument that the property holds.
The second is to use a patched version of the KLEE symbolic execution
engine~\cite{cadar:osdi:2008} that uses Dynamic Symbolic Execution to
individually check all paths through the code.
Finally, using Galois' ``Crux'' verification tool~\cite{crux2020} that can
either perform model checking to to check all paths at once or can perform
symbolic execution on one path at a time.
As we add more verification tools, we will be able to assess the different
strengths and weaknesses of each tool and to give users a choice of which tool
to use for each verification task.

The ability to use the same code as both a test-harness and
a verification-harness is key to achieving our cost-benefit goal: effort
invested in testing helps formal verification and vice-versa; and developers
can continuously tune their productivity by switching between the speed of
running tests and the assurance of running verifiers as they see fit.

The above example has similarities to the proof harnesses used at AWS~\cite{chong:icse:2020}
with one important difference: our approach supports both fuzzing and
verification and supports two verification tools at present whereas their proof
harnesses only support verification and are based on a single verification tool
(CBMC).
We suspect that this difference is mostly one of emphasis and that it would be
relatively easy to make their proof harnesses independent of both the choice of
verifier and the choice between verification and fuzzing.
(It would be very useful to perform experiments like this to understand the
costs of our generalization.)
Work with more technical similarity is DeepState~\cite{goodman:ndss:2018}
which applies a similar approach to C++ development based on the GoogleTest
API.
Their approach is very similar to ours though they put more emphasis on
the ability to change between different verification tools and to change
between fuzzing and verification.

\section{Obstacles and challenges}

The above represents an early experiment testing the hypothesis that it is
possible to achieve a good cost-benefit ratio by integrating elements of formal
verification tools into the developer workflow using testing practices as an
entry point to build on.
Our primary focus at the moment is on
adding Rust support to existing verification tools
and implementing the ``propverify'' library.
While we are not yet in a position to perform a usability evaluation or
to understand how our choices impact verifiability,
even at this early stage there are a range of challenges that need to be addressed if we expect to increase adoption by an order of magnitude.
We break these down into individual usability challenges faced by a developer, ecological challenges and technical challenges.

\subsection{Individual usability challenges}

Based on our anecdotal experience observing developers and researchers as they
apply formal methods and on our previous experience persuading a hardware
design company to adopt formal verification~\cite{reid:cav:2016}, we want to
highlight a number of behaviors that we believe are roadblocks today.

The most pressing issue is that we are asking developers to change their behavior, and expecting behavior change to follow by merely advertising improvements to the weekly cost-benefit
ratio will not be enough. 
The new behavior must be \emph{perceived} to be beneficial by the individuals
who are asked to change their practices: developers need to feel that the
reward, that is the benefits received, exceeds the effort they exerted to
complete the task.
With the developers we interviewed, we observed a degree of skepticism about formal methods, suggesting that they thought the cost-benefit ratio was not likely to be in their favor. Overcoming this skepticism needs to be done by giving them direct positive experiences rather than advocacy, and experiences that can be achieved within the amount of time the developers are willing to risk. This is the primary
motivation behind our hypothesis that the methods must pay for themselves 
within a week.\footnote{If pressed we would say a lot of formal verification of software fails the monthly or even yearly cost-benefit test. So, whilst simple-sounding, this will be a huge challenge.}

We cannot underestimate the influence of developers' existing
beliefs of the utility and role of formal methods. Warranted or not, many
developers do not see how formal methods relate to their work.
We believe we will need to deliberately undertake work to build the relationship
and earn developer trust.
The first step to reconstructing the trust of developers is a supportive onboarding 
experience. Perhaps one that supports developers to form relevant goals
and expectations for using
a technology, install it easily and then to make progress towards the goals they set.
One way of evaluating how we are doing towards this goal is to consider the 
interactions of a developer by applying a simplified version of a cognitive 
walkthrough~\cite{rieman:chi:1995}. Very broadly this entails modelling a user 
engaging with the tool: establishing a goal, considering what options are available 
to them, considering the match between these options and their goal, and evaluating 
the feedback received when they perform the item that matches best. This has been applied in the past to a range of developer activities including consuming APIs and using code completion~\cite{macvean:ppig:2016}.

Existing formal verification technologies have challenges at each of these stages. It is not easy for developers to articulate their goals without understanding what the technology can do, instead the technology is often presented for experts who find it easy to, for example, know why `symbolic execution' might help with fixing bugs.

Even when a goal is established, most linguistic and notational systems like programming languages make it hard to know what options are available. This is not helped by tool designs that focus on expert behavior without much signaling along the way as what is essential for all users to understand and what is only needed in advanced use cases.
For example, the KLEE tool has 140+ possible command line arguments: optimizing
for research flexibility over usability.

Finally, the feedback that developers receive along the way is critical to early success. This applies to both their use of the tools, and to the results they receive, for example, if a bug is detected, knowing where it is within the code is important~\cite{zeller:bugs:2005}. Can the developer read the results and know how to reproduce the defect?

In addition to these fundamental usability challenges, we see evidence that developers experience a lot of difficulty simply getting the tool running. This is not an uncommon problem for tools designed in complex domains. One of the only known ways to simplify installation and setup is for the researchers who design the tools to undertake significant design and engineering work~\cite{macvean:ppig:2016} which unfortunately is often deemed unappealing by researchers even though it is critical and necessary to wider adoption.

This is an especially significant challenge with formal verification technologies as they often require the developer to prepare the program in a specific manner in order to be used with the tool. In some cases it requires compiling third party libraries, which can be challenging in itself. Such issues may appear to be out of scope for formal verification researchers, however, we must find ways to eliminate this complexity rather than asking the developer to deal with it, if we want developers to recognize and appreciate the value of our tools. 
%in order to address this challenge it will be necessary to solve such usability problems that are not formal verification's `fault', but are needed in order for the benefits to become clear.

Once the developers are onboarded and have started using the tool, a different set of usability considerations are necessary to continuing maintaining a positive cost-benefit ratio. We find the Cognitive Dimensions vocabulary to be a helpful articulation of the profile of usability characteristics that will be needed. For a detailed explanation of the vocabulary see~\cite{green:tutorial:1998}.

In order to use verification tools its necessary for the Abstractions to have good Role Expressiveness and Closeness of Mapping. These provide the basis of the bridge between the verification system and the domain the developer is working in, providing a shared description for the developer to decompose their problem into, and the basis on which the verification will report errors. 

The short window in which benefits are returned requires good Progressive Evaluation, in order to know the work is heading in the right direction, this is important to support the developer if they need to put in extra work in, for example, annotating code with loop invariants.

Tools that reason about negative properties (e.g., the program cannot have a null pointer error) have particular additional design needs of low Error Proneness to avoid accidentally configuring the system to miss bugs~\cite{fonseca:ecs:2017} especially in a context where the developer is needing to manage false positives which have a tendency to grow over time~\cite{chou:bugs:2005}.

We suggest that there is work to be done to build a system that meets this design profile, both for onboarding and sustained use. However, by itself even this would not be enough, it would provide a context in which a formal verification was easy to get started with by a single developer, but in order to gain adoption the tools must also be usable within teams and organizations that write the world's software, in other words it must have ecological validity.

\subsection{Ecological challenges}

There are some surface level requirements that need to be met for the tools to function for teams, for example, configuration of the tools needs to be either standardized or committed into a version control system to eliminate potential interference between developers. Just as there are various challenges associated with large scale unit testing, such as management of flaky tests within regression test suites, once there are many thousands of properties being tested~\cite{reid:cav:2016}, there will need to be similar tools for managing the results of verification, turning some tests on and off, tracking over time, etc. Teams will need to agree on the best practices for how they go about using the tools, learnings that we expect to arise from case studies.

However there are also more complex context-specific challenges that may arise
within organizations.
For example, Google and Facebook both found that code review was a good context
in which to report results from static analysis~\cite{sadowski:icse:2015,
sadowski:icse-seip:2018, ohearn:lics:2018}.
It is possible that this might also be a good location for presenting the
results of the formal verification tools.
%
% However other organizations prefer to use pair programming~\cite{pair-programming}
% instead of code review. More research is needed to explore different
% configurations here.
%
Amazon found good results forming a specialized team to provide the interface
between the tooling and the broader development team in order to help ensure
a strong cost-benefit ratio for the developers~\cite{chong:icse:2020}.
However as with the centralized developer tools function at
Google~\cite{Potvin:cacm:2016}, these all work within a single organization and
it is less clear how they can be applied as a model within open source
communities.

These options will need to be clarified in case studies. Probably in the early days focused on
particular programming niches that show the advantages of adding formal verification to a programmer's 
toolkit, and share best practices on the techniques needed to get the most out of the tools
and the pitfalls to avoid. 

Such case studies will also yield valuable empirical data on the cost-benefit ratio. This is one example
of a metric of the effect of introducing formal methods, these metrics create power within organizations~\cite{beer2016metric},
especially ones seeking to use data driven decision making. To support this decision making, it is not sufficient to be useful, the tools must be seen and quantified as being useful, and ideally have their utility predicted ahead of time. However, many formal methods do not work well with existing metrics such as code coverage --- alternatives need to be found to support management decisions about when to and when not to use formal methods, buy-in that will be necessary to achieve order of magnitudes growth in adoption. 

\subsection{A technical challenge: Special-purpose tools; general-purpose engines}

At their core, formal verification tools are based around general-purpose
reasoning engines that can check many different properties about code written
in a given language or that can be compiled to a given intermediate language.
This generality suits developers of tools, but we believe that software engineers
would benefit from more specialized tools focused on a few tasks.

This approach has been applied in the commercial EDA tool space for hardware
development that provides tools specialized for tasks such as
checking the equivalence of two hardware designs;
or
checking for glitches that occur when a part of the design is powered up;
or
checking temporal logic properties of designs.
Hardware engineering teams often build even more specialized tooling around
these tools for tasks such as checking processor pipelines~\cite{reid:cav:2016}.
These different tools are (probably) based on a single formal verification core
based on SAT-solving and model-checking
and could, potentially, be supported by a single general-purpose tool.
The advantage of creating special-purpose tools for each task is that
it allows them to provide interfaces suited to each individual task
to assist in specifying what they want to check,
to insert ``design waivers'' to suppress false-positive reports,
to assist in understanding and localizing reports,
and to generate status reports to assist engineering management teams.

% In the first step of identifying these specialized tasks, we might draw
% inspiration from the different competition categories in the software
% verification competition\footnote{\url{https://sv-comp.sosy-lab.org/2020/}}
% that includes
% %
% checking whether a program can fail by calling an error function;
% %
% checking for memory safety;
% %
% checking concurrency safety;
% %
% checking for overflows;
% %
% and checking for termination.
% %
% Each of these categories is subdivided into the separate tasks of
% finding bugs and showing the absence of bugs
% and, in response to past experience, these categories now require
% the generation of proofs that can be checked by an independent tool.

Creating specialized tools allows us to design specialized interfaces that are
easier to use.
Specialization also enables standardized interfaces that enables users to
switch from one tool to another and enables tool developers to create hybrid
tools that use a variety of different techniques to achieve the goal.
For example, in software verification, we see the rise of ensemble approaches
that run different verification tools in parallel~\cite{beyer:hvc:2017} and in
test generation we see the use of separate techniques such as
fuzzing, symbolic execution and search-based
techniques including genetic algorithms and hill-climbing.
We also see the
emergence of hybrid approaches that combine multiple techniques such as
DeepState~\cite{goodman:ndss:2018} that allows a variety of concolic
execution tools to be used to check C++ properties expressed using
the GoogleTest API
and COASTAL~\cite{visser:tacas:2020} that dynamically alternates between
fuzzing and concolic execution.

We note that the existence of competitions for automatic verification tools has
already driven standardization of the low level API used to build verification
harnesses and annotate programs; and standardization of the result format.
This seems to be an easier task for automatic verification tools than it would
be for an auto-active verification tool such as Dafny~\cite{leino:icse:2013} or
an interactive, general purpose verification tool such as Coq or Isabelle that
have more interface to standardize and
require more knowledge of how the tool works.

Creating specialized tools for particular tasks is, unfortunately, in tension
with another important goal:
creating an integrated toolflow where artifacts (such as test harnesses),
annotations, etc.  can be reused for multiple purposes.
Such reuse is critical to achieving our weekly cost-benefit ratio goal.
To allow reuse we must avoid fragmentation between tools specialized for
different verification tasks.
Finding the right compromise between different tasks will likely need multiple design iterations.

\subsection{Predictability vs. decidability}

It is not yet clear how to handle the conflict between power and decidability of verification
problems.
Below what Leino calls the “decidability ceiling”~\cite{leino:informatics:2001}, we would hope
that most tools can consistently prove that properties hold.
But, as we use more properties outside of the decidable subset of their
property language (e.g., as non-linear arithmetic is used), we can expect that
tools will diverge: some succeeding, some taking a long time but
eventually succeeding and some failing to prove the property.
We might also expect that different versions of the same tool or different runs
of the same tool will have different outcomes.
Using an ensemble of different tools to build on the strengths and weaknesses
of individual tools~\cite{beyer:hvc:2017} will help to some degree but,
fundamentally, attempting to prove overly hard properties
will behave like (and be as unsettling as) having flaky tests
in a testsuite.
To deal with this, verification tools need to support profiling
of code to find code that is impacting verification time~\cite{bornholt:oopsla:2018}
and we need to develop guidelines to support what the hardware industry calls
``Design for Verification'': changing the design to enable/simplify verification.

% \subsection{Lessons from bug-finding tools}

% False positives (i.e., false bug reports) can be a major problem because they
% decrease users trust and value in a tool.
% %
% This issue is discussed in Godefroid's provocative
% argument~\cite{godefroid:bugs:2005} that ``the soundness of bugs is what
% matters'' and in Cok's recommendations~\cite{cok:bugs:2005} for successful
% deployment of bug finding tools.
% %
% Worse, long term use of a bug-finding tool such as Coverity leads to an
% accumulation of false positives in a codebase because developers are able and
% motivated to fix bugs but false positives can, at best, be worked
% around~\cite{chou:bugs:2005}.

\section{Conclusion}

We aim to increase the uptake and usability of formal methods by developing
libraries and tools that build on activities that programmers are
already comfortable and effective with (principally testing) and give some of
the benefits associated with formal verification.
Our work is at a very early stage at present and, while we think that this is
a productive direction, it is not yet clear how close we can get to achieving
conventional formal verification goals.

This is a challenging space to work in. `Just' the process of building an 
interdisciplinary team that can talk in the same language about the usability,
social, technical and engineering aspects of the project has been an 
interesting research activity.

In many ways formal verification and testing represent two different modes of engagement with the world. Formal verification sees a model  as a logically coherent entity, and once the world has been suitably translated into that model, it can be rationally and productively reasoned about as a whole. Testing, on the other hand, views the world as being made up of a series of empirically observed points in behavioral space with an ---mostly implicit--- assumption that those empirically observed points generalize.

On the performative spectrum between the logician and the scientist, current software engineering is unusual in how far over towards the logician it starts, simply because of the use of notation and linguistics to define the behavior of programs as a series of steps and rules, rather than a series of examples.
It is not surprising given this that there is a strong instinct to pursue this route with the introduction of even stronger logics to fix the problems that arise when software meets the world. However there is a danger here that testing points out, be it software testing, or the testing that happens when software gets put in the hands of the users. Empirical reality fights backs against the model~\cite{smith:book:1996}. Learning from this resistance as early as possible helps.

In this paper we have tried to turn this problem upside down, by starting from where developers are today, focusing on what they are already doing. We have seen this approach in other areas of computing where previously expert tasks need to be adopted by a wider audience (e.g., in the early 2000s the usable security community grappled with what it means to do human-centred work suggesting that useful, secure applications may be more readily adopted than usable security tools~\cite{smetters:nspw:2002}). 
In a similar spirit, we urge the community to begin from the empirical reality they and their customers inhabit, then deliberately consider what it would take to integrate formal method tooling into existing practice, with an eye toward establishing an attention-economic deal of weekly benefit that they might accept.

We believe that there is not much choice here if we seek broad adoption: to achieve the
benefits of the tools of formal reasoning we must start from
the messy contexts of tests, and `as lived' tools and build out from there.

% \begin{acks}
% \end{acks}

\bibliographystyle{ACM-Reference-Format}
\bibliography{RelatedWork,misc}

\end{document}